\documentclass[11pt, oneside]{article}   	
\usepackage{geometry}                		
\usepackage{graphicx}				

\usepackage{amsmath}								
\usepackage{amssymb}
\usepackage{authblk}
\usepackage{natbib}
\newcommand\Tstrut{\rule{0pt}{2.9ex}} 

\title{Finding Scientific Communities In Citation Graphs:Convergent Clustering}
\author[1]{Shreya Chandrasekharan}
\author[2]{Mariam Zaka}
\author[2]{Stephen Gallo}
\author[3]{Tandy Warnow}
\author[1]{George Chacko}
\affil[1]{Netelabs, NET ESolutions Corporation (an NTT DATA Company), McLean, VA, USA}
\affil[2]{American Institute of Biological Sciences, Herndon, VA, USA}
\affil[3]{Department of Computer Science, University of Illinois at Urbana-Champaign, Urbana, IL, USA}

\providecommand{\keywords}[1]{\textbf{\textit{Index terms---}} #1}
\begin{document}
\maketitle


\begin{abstract} Understanding the nature and organization of scientific communities is of broad interest.  
The `Invisible College' is a historical metaphor for one such type of community and the search for such `colleges' can be framed as the detection and analysis of small groups of scientists working on problems of common interests. Case studies have previously been conducted on individual communities with respect to their scientific and social behavior. In this study, we introduce, a new and scalable community finding approach. Supplemented by expert assessment, we use the convergence of two different clustering methods to select article clusters generated from over two million articles from the field of immunology spanning an eleven year period with relevant cluster quality indicators for evaluation. Finally, we identify author communities defined by these clusters. A sample of the article clusters produced by this pipeline was reviewed by experts, and shows strong thematic relatedness, suggesting that the inferred author communities may represent valid communities of practice. These findings  suggest that such convergent approaches may be useful in the future.

\end{abstract}

\section*{Introduction} 

In this article, we report on an effort to use citation data to identify groups of scientific articles that may reflect widespread small-scale organization in the scientific enterprise. 
We are inspired by the `Invisible College' concept, which appears to originate from a group of intellectually active individuals who held meetings around 1660 and became the 
Royal Society of London in 1663~\citep{price_1966,royal_society}  but more generally refers to a relatively small self-assembled group of scientists with common scientific interests. 

Importantly, there is a sense that these colleges are `in groups' with influence over prestige, research funding, and the scientific ideas of their community~\citep{price_1966}. 
Thus, these groups may also advocate for or exhibit resistance 
to new ideas within their domains of interest~\citep{barber_1961}. 
Furthermore, while such groups may espouse idealized norms of science~\citep{merton_1957}, they are also likely to be driven by social interests such as personal recognition that influence both individual and collective behavior~\citep{barber_1962, crane_1972, hagstrom_1965}.  

An important distinction has also been made between local (small) and global groups in the discussion of community~\citep[p.~112]{hull_1988}, with small groups being credited as the locus of rapid change and innovation.
Furthermore,   Price  noted, in apparent reference to Invisible Colleges, that communications are likely challenged in groups larger than 100 members~\citep{price_little_1963} and that the small `strips' at the research front of science may be at most the work of `a few hundred' persons~\citep{price_1965}; this number  is also cited in a clustering study \citep{Small1985}. 

In a case study \citep{price_1966} of the Information Exchange Group No 1 \citep{green_1965} that was organized by the US National Institutes of Health to focus on electron transfer and oxidative phosphorylation,  
Price and Beaver described a group of 517 members, 62\% of whom were from the United States and the rest from 27 different countries. Price and Beaver used the memos shared within Information Exchange Group No 1 as proxies for research articles and citations. They noted 1,239 authorships in 533 memos with two-author memos being the mode that was stable across a 5 year period. The majority of these authors were  were associated only with a single memo, and the top 30 authors each contributed to six or more memos.  
Three conclusions were drawn from this study: first, that there existed a small nucleus of highly active researchers and others who collaborated with them only once; second, that separate groups existed within this college, and third,  that collaboration was a key feature. This valuable case study of Price and Beaver is, however, limited by examining a tiny sample of the enterprise as it existed in the first half of the decade of 1960-1970.  It is far more likely that a range of group sizes and behaviors exists now, and perhaps even then.

Thus, a natural question is how these small groups form and grow, and how to detect them
from  the bibliographic literature, which our study addresses.
Our study is motivated by two considerations. 
First, the scientific enterprise has grown considerably, experienced greater globalization~\citep{wagner_2008} in the 21st century, and exhibits new features like large international collaborations such as the Human Genome Project~\citep{hgp_2001} and the Advanced LIGO project~\citep{Harry2010}. Even so, the tendency of scientists to form small groups that collaborate to advance their scientific and social interests is unlikely to have vanished and these groups may well exist within larger structures. Thus, we seek to understand organizational structures in the modern scientific enterprise and reconcile them with the observations of Price and others from the 1960s. 
 Second, modern bibliographies and accessible computing make it possible to revisit the search for these small  communities of practice (and in particular invisible colleges)  at scale through community finding analyses of citation graphs.

In our use of citation data to identify and characterize putative colleges or communities of practice~\citep{cop_1991}, our working hypothesis is that such groups can be detected by identifying clusters of articles that are citation-dense, since common interests will result in citation of relevant documents especially those authored by the `in group'. Furthermore, as we noted earlier, we are particularly interested in small communities, since these may represent new ideas~\citep[p.~112]{hull_1988}. 

Rather than attempting to directly identify author communities of practice, we first construct article clusters, and then examine the authors within the article clusters; the motivation for this approach is that 
each community of  practice is by its nature formed around a specific research question or area, 
while individual scientists may participate in multiple communities of practice based on different scientific and social interests. 

\begin{figure}[h!]
\begin{center}
\includegraphics[width=12cm]{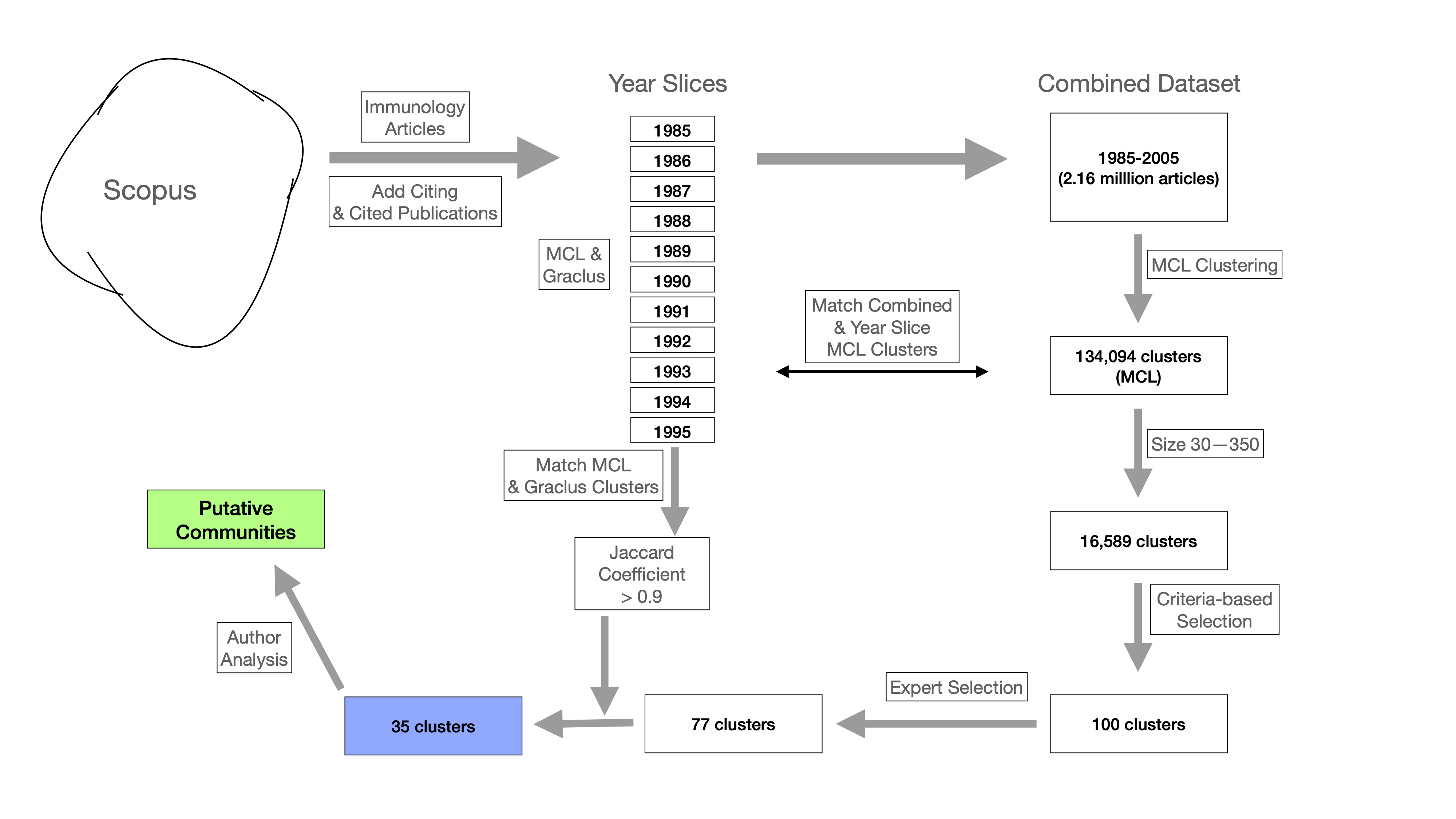}
\end{center}
\caption{Workflow to detect communities. Year-slices (center of schematic) were generated from immunology articles in Scopus for the years 1985-1995 (Materials and Methods). 
The citation data in these year-slices were de-duplicated and combined to construct the combined dataset (right side of schematic). MCL clustering was performed on each year-slice as well as the combined dataset. This produced a  set of 134.094 clusters. We then restricted attention to  those clusters containing between 30 and 350 publications,  which resulted in 16,909 clusters. From these, a sample of 100 clusters was provided to 
two evaluators who rated 77 of them as strongly themed. 
In parallel, each of the 11 year-slices was clustered with MCL and with Graclus. Every cluster from the set of 16,909 was matched, using overlap as matching criterion, to a single MCL cluster from all clusters generated in the 11 year-slices. Each of these matched MCL clusters from the year-slices was also matched to a Graclus cluster from the same year. Convergence was measured using the intersection/union ratio (Jaccard Coefficient) between members of a pair of clusters. 
The 77 clusters selected by the evaluators were further constrained to those with a Jaccard Coefficient greater than 0.9 from their corresponding MCL-Graclus pair; this constraint resulted in 35 clusters. Authors for the publications in each of these clusters were analyzed for the purpose of community inference.} 
\label{fig:pipeline}
\end{figure}

We begin by identifying article clusters constructed from direct citations of articles. The use  of article-level citation patterns is motivated by the understanding that clusters defined by articles that cite each other can be more informative than clusters derived from journal-derived categories~\citep{Waltman2012,Shu2019,Milojevic2019} or from searches for words or phrases that recur in articles~\citep{Klavans2017}. We further restrict our examination to moderately-sized clusters based on the realization that  article clusters that are either very small or very large will not assist in discovering author clusters of interest. 
Our pipeline (shown in Figure \ref{fig:pipeline}) shows the steps we used in this development of
author clusters.

We have chosen the field of immunology as a test case since it has existed for many years, grown and diversified over time, and exchanges discovery and methods with other biomedical areas. To identify article clusters,  we first constructed a citation graph for publications in the field of immunology within an 11 year timeframe, consisting of 2.16 million articles. 
We  then used two different clustering methods, Markov Clustering \citep{dongen2000cluster}  and Graclus \citep{Dhillon2007}, which we previously used to cluster citation data within a discipline~\citep{devarakonda_2_2020}, to identify clusters that are robust to choice of clustering method.
To evaluate the approach,  we use expert annotation on a sample of the  Markov Clustering (MCL) clusters we produced that passed several stringent quality criteria (Fig.~\ref{fig:pipeline}).

These two clustering methods use different strategies and criteria. 
Markov Clustering (MCL) has several desirable features:  it  is scalable, does not require pre-specification of the number of clusters to be generated, and has tunable parameters that control breadth of search and granularity of output. 
Graclus is a spectral clustering method that we have previously used to construct article clusters from citation data~\citep{devarakonda_2_2020}. 
In order to identify those clusters of interest, we limit our attention to those MCL clusters that have very high overlap with at least one cluster produced by Graclus: 
the convergence of these two methods serves to identify those publication clusters where the citation signal is high enough to result in robustness to the choice of clustering method.
By restricting the set of clusters to a range of sizes consistent with a potential community of practice, we are able to 
select publication clusters of interest that are then used to build author clusters that may represent communities of interest.

We explore the merit of this  combined approach by examining the publication clusters it produces, and use human experts to evaluate a sample of the selected clusters for thematic relatedness. 
Our study shows strong concordance between the cluster conductance and expert evaluation of thematic relatedness, supporting the potential value of this protocol.
Thus, we consider this study a first step in designing and testing a pipeline that could enable large-scale identification of communities of different sizes and types, based on different search criteria. choice of clustering algorithm, and parameters.

\section*{Materials and Methods} 

As a source of bibliographic data for this study, we used Scopus~\citep{scopus_ref}, as implemented in the ERNIE project~\citep{Korobskiy2019}. At the time of this analysis, our Scopus data consisted of $\sim$95 million publication records plus their cited references. 
From these data, we selected publications in English, of type `article' with publication type `core', and Scopus All Science Journal Classification (ASJC) code of 2403, which corresponds to immunology, for each of the years 1985--1995. 
We then  amplified the set of immunology articles thus extracted by supplementing them with articles that directly cited them and as well as by articles cited by these immunology articles. 
The only constraints we imposed on the  cited or citing articles were to require that they were English publications of type `core'; in particular, the cited and citing articles were not constrained by ASJC codes.
All in all, we assembled 12 datasets (Table \ref{tab:empirical-MCL}) that consisted of (i) 11 `year-slices', each representing the set of immunology articles published in a given year (any of 1985-1995) along with their cited references and the publications 
that cited them (e.g., the 1985 year-slice) and (ii) the union of data from the 11 year-slices, to create a working dataset of  2,163,683 articles that we refer to as the combined dataset. These data were stored as an annotated list of 6,846,323 pairs representing 11 years of data.

For Markov Clustering analysis, we downloaded and compiled source code for the MCL-edge software~\citep{dongen2000cluster}. After evaluating different runtime parameters, we clustered test sets using an expansion parameter of 2 (default) and an inflation parameter of 2.0 to minimize 
the number of large aggregated clusters (Figure \ref{fig:fig2}). We generated clusters using the same parameters for each of the 11 individual years.  See Table \ref{tab:empirical-MCL} for empirical properties of the MCL clusters computed for the 11 year-slices and the combined dataset.

Under the same conditions, we also  generated 134,094 clusters containing 2.16 million nodes from our working dataset (above), resulting in cluster sizes of 1 (minimum), 16.2 (mean), 9 (median), and 3,956 (maximum).  
For a random graph comparison, we performed 1 million reciprocal citation exchanges between randomly selected pairs of publications on these data and then ran MCL-edge on the resultant data. 

To evaluate clusters and shuffled-citation clusters generated by  MCL, we measured (i) cluster conductance~\citep{Shun2016,Emmons2016,devarakonda_2_2020}, which essentially measures intra-cluster density, 
and (ii) textual coherence using the Jensen-Shannon divergence~\citep{Boyack2011},  with the expectation that `good' clusters  would exhibit low conductance and high coherence, respectively.  
We also clustered data from each of the 11 years using Graclus, adjusting its runtime parameter number of clusters to produce a distribution of  cluster sizes that roughly approximated those generated by MCL (Figure \ref{fig:fig2}).

\begin{table}
\begin{center}
\scalebox{0.9}{
\begin{tabular}{ccccccc}
  \hline
Dataset & num\_clusters & num\_articles & mean\_size & median\_size & mean\_cond. & mean\_coh.\\ 
  \hline
1985 & 10,568 & 293,086 & 27.73 & 	18 &	0.30 & 0.09 \Tstrut\\ 
1986 & 10,621 & 310,030 & 29.19 & 	19 &	0.32 & 0.09 \\ 
1987 & 10,984 & 325,661 & 29.65 & 	20 &	0.31 & 0.09 \\ 
1988 & 11,697 & 363,038 & 31.04 & 	21 &	0.33 & 0.09 \\ 
1989 & 12,401 & 397,292 & 32.04 & 	21 &	0.33 & 0.09 \\ 
1990 & 12,542 & 419,500 & 33.45 & 	23 &	0.34 & 0.09 \\ 
1991 & 13,089 & 463,581 & 35.42 & 	24 &	0.34 & 0.09 \\ 
1992 & 13,878 & 507,365 & 36.56 & 	25 &	0.35 & 0.09 \\ 
1993 & 14,135 & 542,948 & 38.41 & 	27 &	0.36 & 0.09 \\ 
1994 & 14,681 & 584,768 & 39.83 & 	27 &	0.38 & 0.09 \\ 
1995 & 15,918 & 642,686 & 40.37 & 	28 &	0.37 & 0.09 \\ 
combined & 134,094 & 2163683 & 16.14 & 9 &	0.70 & 0.09 \\ 
   \hline
\end{tabular}}
\caption{Empirical properties of the MCL  clusters computed for each year-slice as well as for the combined Immunology dataset.  We show statistics regarding size, conductance (cond.), and coherence (coh.) }
\label{tab:empirical-MCL}
\end{center}
\end{table}

To compute textual coherence, we used the titles and abstracts of all the articles in our study. On average, roughly 11\% of the publications had missing titles and/or abstracts, reducing the corpus to titles and abstracts from 1.95 million (1,955,164) articles. We first concatenated these titles and abstracts and pre-processed them by lemmatization based on parts of speech (POS) tagging, to preserve the tokens and their context. 
The four POS considered were adjectives, nouns, adverbs, and verbs. 
For all other types of parts of speech (including those not classified at all), the token was mapped to `noun' by default;
for example, `grow' and `growth' are two different tokens while `grow' and `growing' would be mapped to `grow'. 
We then removed stop-words using a  list of 513 tokens comprising basic NLTK stop-words, 
PubMed stop-words, and a select list of tokens from the top 500 most frequent words in our dataset.

For each cluster of size greater than 10 (after removing missing values), we performed a second pre-processing by filtering out those tokens that occurred only once in the entire cluster. Next, we converted all the remaining tokens by article in the cluster into a matrix of term frequencies (i.e., for each article, we had a vector of counts of all the tokens). 
We also obtained a vector of counts for all the unique tokens in the cluster. Textual coherence was measured by using the Jensen-Shannon Divergence (JSD), which is used to compute the distance between two probability distributions. JSD was computed between the vector of term frequencies of the cluster and each article in the cluster using the following:
\begin{equation*}
JSD_{p,q} = \frac{1}{2}D_{KL}(p,m) + \frac{1}{2}D_{KL}(q,m)
\end{equation*}
where $m = \frac{p+q}{2}$, $p$ is the probability of a term in a document, $q$ is the probability of the same term in the cluster, and $D_{KL}$ is the Kullback-Leibler divergence, given by:
\begin{equation*}
D_{KL}(p,m) = \sum{p_{i}log(\frac{p_i}{m_i})}
\end{equation*}

We computed the textual coherence for a given cluster $X$ of size $n$ (after removing missing values) as follows.
Letting JSD\textsubscript{X} denote the arithmetic mean of all article JSD values in $X$, we define the textual coherence of $X$ to be JSD\textsubscript{X}-JSD\textsubscript{random(n)}~\citep{Boyack2011},
where JSD\textsubscript{random(n)} denotes the JSD of a  random cluster of the size $n$.  

JSD\textsubscript{random(n)}  is the arithmetic mean of all the JSD values computed from  random selected sets of size $n$  from all the articles in our study. 
For each value $n$, we estimate JSD\textsubscript{random(n)} by selecting 50 article subsets of size $n$ at random, and averaging the results.
The method of computing each iteration of JSD\textsubscript{random(n)} is exactly the same as the method described for JSD\textsubscript{X} above.

\begin{figure}[h!]
\begin{center}
\includegraphics[width=7cm]{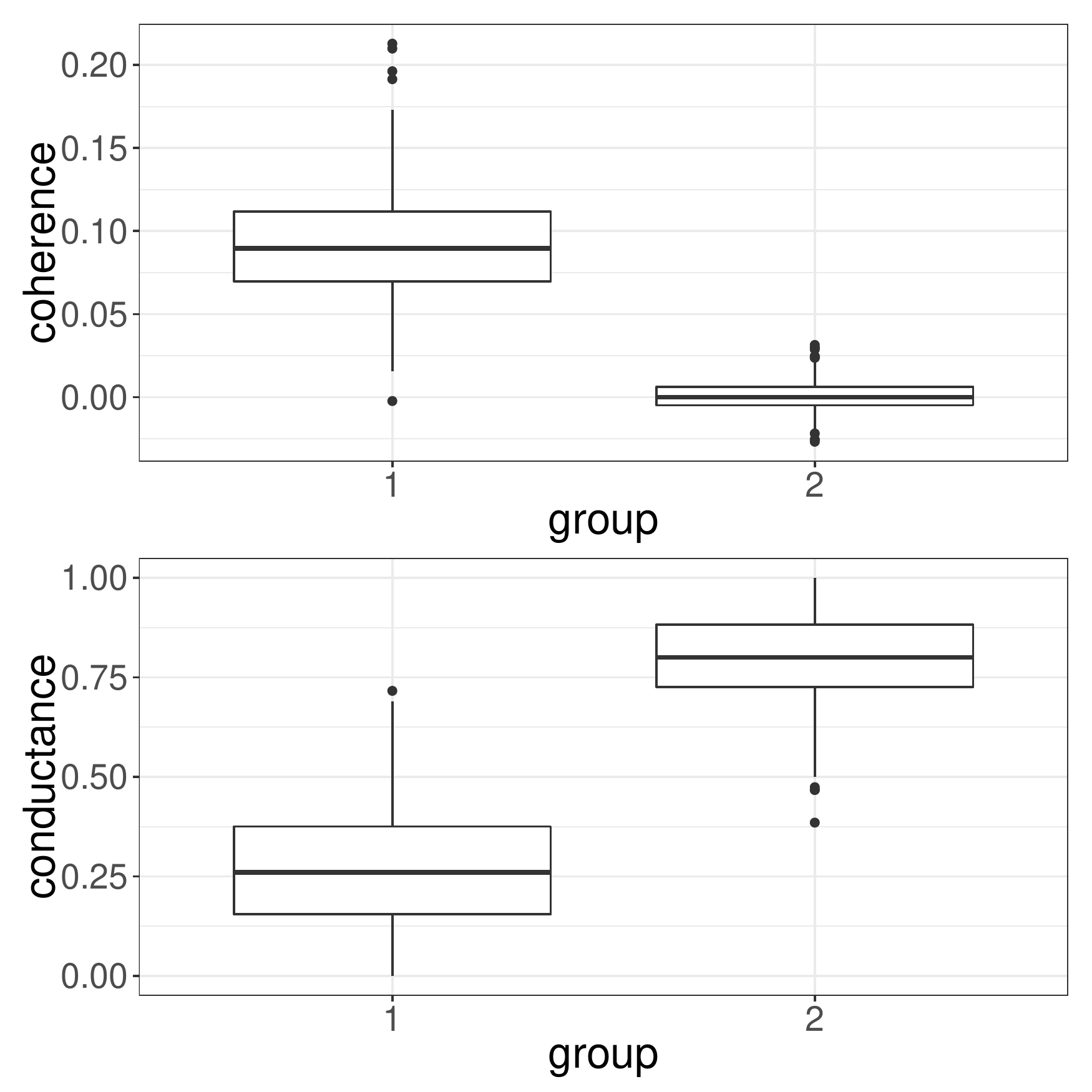}
\end{center}
\caption{Conductance and coherence profiles of 1,000 MCL clusters  (group 1) compared to random clusters (group 2), showing that MCL clusters have lower conductance and increased coherence compared to random clusters of the same size. The 1990 immunology year-slice was either clustered (x-axis: group 1) or subjected to 1 million citation shuffling (group 2) operations and then clustered using MCL-edge software with an expansion parameter setting of 2 and inflation parameter setting of 2.0. From each of the resultant datasets, a sample of 1,000 clusters of size 30--350 publications were randomly selected and analyzed for conductance and coherence.} 
\label{fig:cond-coh-mcl}
\end{figure}

\begin{figure}[h!]
\begin{center}
\includegraphics[width=10cm]{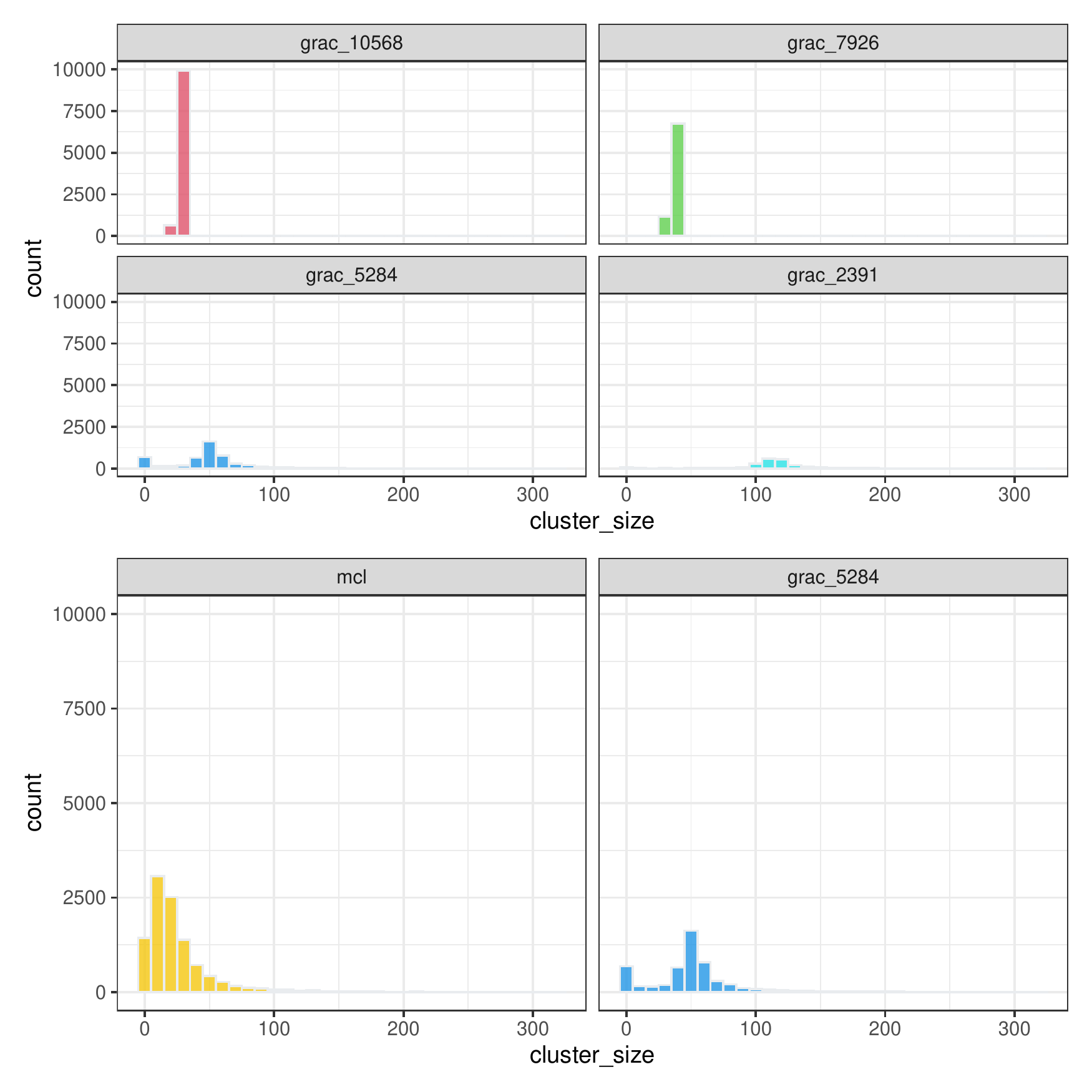}
\end{center}
\caption{Cluster Size Distributions for MCL and Graclus. Citation data were clustered using either Markov Clustering (mcl) or Graclus (grac\_x where x is the number of clusters specified when running Graclus).  Upper panel (four sub-plots): distribution of cluster sizes from Graclus runs with the number of clusters specified in each run varying across (2391, 5284, 7926, 10,568). Lower panel: Running Graclus while specifying roughly half the number of clusters generated by MCL resulted in a comparable distribution of cluster sizes (x-axis), which were then examined for convergence of size and content across the two clustering methods. Data are shown for the 1985 year-slice, which contains 293,086 publications and formed 10,568 clusters by Markov Clustering.} 
\label{fig:fig2}
\end{figure}

After completion of MCL clustering and computing conductance and coherence values, we compared a random sample of 1,000 clusters from the year 1990 and visualized the effect of shuffling citations on conductance and coherence compared to the original citation data (Figure \ref{fig:cond-coh-mcl}).

We scored each publication using a weighted citation count (Williams et al 2015, Keserci 2017) of intra-graph citations, which assigns a score to each node in a graph that consists of the number of in-graph citations it receives plus the number of citations received by its neighbors. We also computed the number of times each article in our working dataset had been cited in Scopus between publication and  July 2020.

\emph{Thematic relatedness}. Lastly, we used the expertise of two of the authors of this article (Zaka and Gallo), who are professional peer review specialists in the biomedical sciences, and highly experienced at  clustering proposals for funding based on multiple criteria such as sub-disciplines, methods, disease, and researchers.
In preliminary experiments, we provided a small number of training clusters to these evaluators representing a range of conductance values to assist in develop a common set of principles by which they would evaluate a test set. The clusters in this development set were not considered further.

For expert evaluation, we randomly selected 90 MCL clusters, each with 30--350 publications, and  with conductance values of no more than 0.5. 
Since smaller clusters occur more frequently, the sample of 90  was constructed from two strata based on size to ensure representation of the larger cluster sizes.  An additional 10 clusters with conductance values greater than 0.5 were added to the sample. 
The two evaluators were  each asked to evaluate 50  selected clusters (45 from the set of 90 and 5 from the set of 10) for thematic relatedness, given only the titles and abstracts for each publication in each cluster. 
Using their expertise in peer review, they assigned scores on a simple scale of 1--4 where 1 represented a cluster exhibiting a single discernible scientific theme, 2 for a moderate level of thematic relatedness, 3 for poor thematic relatedness, and 4 for `unable to evaluate'.
The evaluators also annotated each cluster with keywords such as `hemophilia' or `adenosine deaminase' to indicate the theme that they discerned (Supplementary Data). 

\section*{Results \& Discussion} 

\emph{Markov Clustering of 2.16 million publications}. Our initial experiment was to cluster the 2.16 million publications in our combined dataset as well as separately for the 11 individual year-slices using Markov Clustering (MCL). 
This experiment resulted in 134,094 clusters for the combined dataset and from 10,000-16,000 clusters for each of the individual year-slices,  as shown in Table~\ref{tab:empirical-MCL}.  

Some noteworthy trends are immediately apparent.
First, the number of publications  increases monotonically each year between 1985 and 1995 and the number of clusters generated by MCL (as well as the average and median size of these clusters) similarly increases for each year. 
However, while the number of clusters computed on the combined dataset is roughly ten-fold the number of clusters in individual years,  the average  and median cluster size decrease by roughly two-fold for the combined dataset (Table~\ref{tab:empirical-MCL}).

Another interesting feature is that except for the combined dataset where the mean conductance is somewhat high (0.70), the MCL clusters have very good coherence values (mean of 0.09) and  conductance values (mean of 0.30-0.38).
However, the conductance values are  of greater interest here than the coherence values, since by measuring citations they more directly assess the likely
interactions between authors.

A comparison of conductance and coherence profiles of 1000 MCL clusters from the year 1990 to random sets of publications of the same size is shown in Figure  \ref{fig:cond-coh-mcl} (Materials and Methods).
This comparison shows that random subsets of publications have much poorer conductance and coherence values, highlighting that MCL clusters are of very high quality with respect to both criteria.

\begin{figure}[h!]
\begin{center}
\includegraphics[width=10cm]{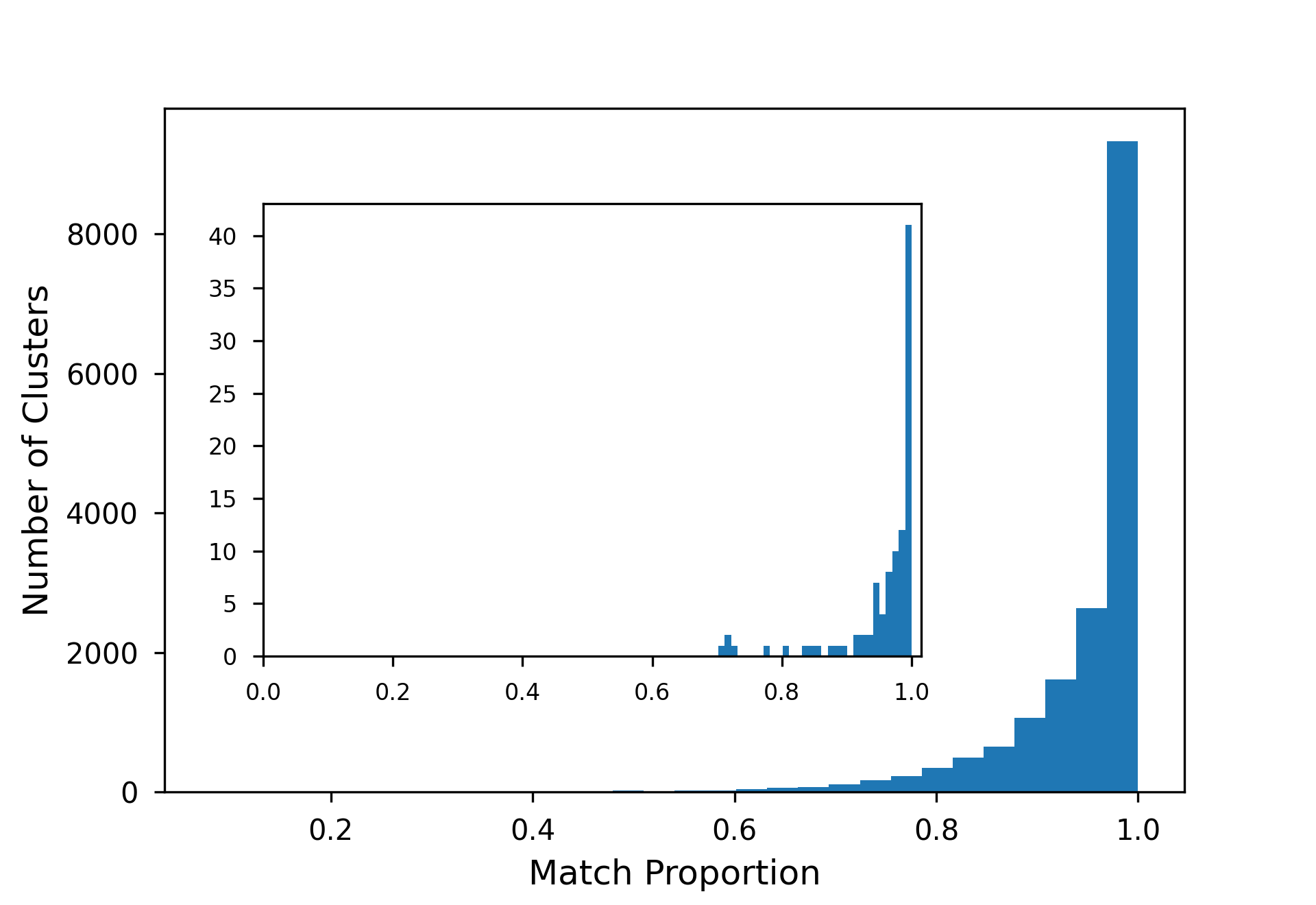}
\end{center}
\caption{Combined dataset to year-slice matching. The histogram shows the proportion of each of 16,909 MCL clusters of size 30--350 that is found in the single best-match MCL cluster from the 11 year-slices. Values ranged from 0.081 to 1.0 with an average of 0.946. Inset: the same measurements are shown for the 100 MCL clusters that were evaluated by humans. } 
\label{fig:fig4}
\end{figure}

We also observed that each MCL cluster in the combined dataset mapped well to a single MCL cluster in some year-slice but not as well to any MCL cluster in another year (Figure \ref{fig:fig4}).  As an example, cluster \#3780 from the combined dataset is a set of 70 publications focused on the enzyme adenosine deaminase; mutations in the gene encoding adenosine deaminase cause a severe combined immunodeficiency (SCID) phenotype. This cluster  matches best to cluster \#922 consisting of 103 publications from the 1995 year-slice with 84.3\% the first cluster (\#3780) drawn from the combined dataset found in cluster \#103 from 1995. 
In comparison, the proportion of the first cluster found in the ten remaining year-slices ranged from 7.1\% to 55.7\%. 
A second example provided is cluster \#122 from the combined dataset, concerning hemophilia and consisting of 134 publications; this cluster from the combined dataset matched best to cluster \#290 from 1993 of size 168 with 99.3\% of the first cluster (\#122) contained in cluster \#290  from 1993.
 In comparison again, the proportion of cluster \#122 in the best matches to clusters from the remaining ten year-slices ranged from 0\% to 28\%.  
 This observation turns out to be useful in enabling a direct comparison between MCL clusters and Graclus clusters, as we now describe.

Graclus requires that the number of clusters be provided as a runtime parameter.  We found that setting this parameter to half as many clusters as those generated by MCL for a given year-slice resulted in a distribution of sizes that overlapped with MCL (Figure~\ref{fig:fig2}). 
However, Graclus did not run to completion on the combined dataset when we set the runtime parameter to 67,000 clusters, or even when we set it to 30,000 clusters. 
Therefore, we developed a technique to pair each MCL cluster to a Graclus cluster that maximized the overlap with the MCL cluster, computed as follows.
Specifically, we used the Jaccard Coefficient, which is the ratio between the size of the intersection and the size of the union of the two sets; this is maximized at $1.0$ (when the sets are identical), and is $0.0$ if the two sets are disjoint.  
By restricting our attention to those MCL clusters that have Jaccard Coefficient greater than 0.9 to their paired Graclus cluster, we were able to  identify a subset of the MCL clusters that were also (nearly perfectly) recovered in the associated Graclus cluster, and so represent clusters that are less likely to be an artifact of the choice of clustering method.
This technique also specifically associates each MCL cluster with its nearest Graclus cluster, and so enables a comparison between MCL and Graclus clusters.
For example, an influenza virus MCL cluster (\#1189) from the combined dataset mapped best to MCL cluster \#116 from the 1988 year-slice. This cluster (\#116) from 1988, with 190 publications, matched Graclus cluster \#3164 with 194 publications. 
The Jaccard Coefficient (overlap) between the two clusters was very high, equal to 0.96
(Table \ref{tab:tab3}, Row 1).

We further restricted our attention to that subset of the MCL clusters that  were  rated 1 by the evaluators. 
Of the 100 clusters given to the two experts to evaluate (Materials and Methods), 77 clusters were rated 1, 18 clusters were rated 2, and 5 clusters were rated 3 (no clusters were rated 4), suggesting that the evaluators considered roughly three quarters of the clusters to be strongly themed given their knowledge and experience.
35 of these 77 MCL clusters also had Jaccard Coefficients greater than 0.9, and so were selected for further study. 
These MCL clusters, 35 in all, exhibited very low conductance values of 0.01--0.17 (and in fact 32 of the 35 had conductance values of at most 0.1), suggesting exceptionally good clustering quality (Table \ref{tab:tab3}).
Interestingly, the coherence values range substantially across the 35 clusters, from as low as 0.04 to as high as 0.14, and while the higher end of this range is  good, even the top end of the range is not exceptionally good.
In general, as expected, coherence values are not as good a predictor of cluster quality as conductance.
  
\emph{Comparing MCL and Graclus clusters. }
Our study enables a comparison of MCL and Graclus clusters, as performed on this dataset.
 Interestingly, despite heavy overlap, the conductance values of MCL clusters often differed from those of Graclus clusters,
 with MCL clusters usually exhibiting low conductance values (median of 0.032) and Graclus clusters usually exhibiting high conductance (median of 1.0, 25 of 35 Graclus clusters had a conductance of 1.0). We traced this observation to nodes of high degree not being present in matched Graclus clusters.

For example, cluster \#198 of size 325 from the combined dataset mapped to an MCL cluster of size 328 and a Graclus cluster of size 327. 
The Jaccard Coefficient for these two clusters is 0.99.
However, the two clusters differ substantially in conductance:  the MCL cluster has conductance of 0.01 whereas the conductance of the Graclus cluster is 1, which can be traced to the absence of any internal edges in the Graclus cluster (Table \ref{tab:tab3}, Row 2, Column: int.edges(g)). 
Although this is a general trend, in some cases both clusters have good conductance values.
For example, cluster \#1595 of size 111 from the combined dataset mapped to an MCL cluster of size 114 and a Graclus cluster of size 112 from the 1993 year-slice. The Jaccard Coefficient for these clusters was 0.9825, and
both the MCL and Graclus clusters have very low conductance of  0.01 (Table \ref{tab:tab3}, Row 3).

Overall  for the 16,909 pairs of clusters (derived from restricting 134,089 MCL clusters in the combined dataset to  those of size 30--350), the conductance of MCL clusters was better than those of the corresponding Graclus clusters in $\sim$59.5\% of the cases, MCL and Graclus conductance were equal in $\sim$5\% of the cases, 
and MCL conductance was worse than Graclus conductance in $\sim$35.5\% of the cases. 
When a restriction of greater than 0.9 is placed on the Jaccard Coefficient of paired MCL and Graclus clusters (3,669 cases), the conductance of MCL clusters was better than those of corresponding Graclus clusters in $\sim$50.5\%, equal in $\sim$18.3\%, and MCL conductance was worse than Graclus conductance in $\sim$31.2\% of 3,669 pairs.

At an aggregate level, the median conductance of all  16,909 MCL clusters  was 0.165 versus 0.219 for the corresponding Graclus clusters. 
When this set of 16,909  MCL clusters was reduced to those with  a Jaccard Coefficient of greater than 0.9, the median conductance for MCL was 0.072 versus 0.219 for their paired Graclus clusters. 
Finally, only 6 of 16,909 MCL clusters had a conductance value of 1 ($\sim$0.04\%) compared to 4,545 Graclus clusters ($\sim$26.88\%). 
 Thus, MCL clusters tend to have lower (i.e., better) conductance values compared to matched Graclus clusters. 
 
In contrast, a comparison of coherence values for the paired MCL and Graclus clusters we evaluated shows very little difference between the two methods. 
This is not surprising since the overlap between matched clusters is high (Jaccard Coefficient greater than 0.9), and coherence measures average textual similarity between members of the cluster and an overall representation of the cluster.
The significance of coherence values in interpreting cluster quality, however, is questionable for this context, where we are using article clustering in order to detect author communities; that is, at the best, high coherence would indicate that papers are examining similar questions, but would not indicate that the authors of the paper are even aware of each other, while citation-based clustering metrics, such as conductance, avoid this issue. 
We also have reduced interest in  coherence values for these data, due to the average loss of 11\% of titles and abstracts across clusters (Materials and Methods).

We have found cases where further analysis of selected clusters  (and those not selected) will be necessary in order to identify author clusters, either using external edges, in-graph citation counts, or in-Scopus citation counts. 
We present three edge-cases (considering intra-cluster edges only)  to illustrate this point.

Cluster \#117927 contains a single article \citep{White1989} that is focused on Staphylococcal Enterotoxin B; it has 351 external edges from other clusters in the 
set of 134,094. An alternate perspective is that this cluster could be reconfigured to consist of itself and all its citing nodes to form its own community. 

Cluster \#1301, focused on 
non-steroidal anti-inflammatory drugs (NSAIDS), consists of 125 publications, and 123 of its articles are cited by a single article from the cluster~\citep{Cucala1987}, which is cited, in turn, 
by another article in the cluster. 

Finally, Cluster \#1673 consists of 108 publications, one of which~\citep{michel_1992}, a comparison of hypersensitivity vasculitis and the Henoch:Schonlein purpura, is cited by the 
other 107. 

In comparison, Cluster \#559 consists of 185 publications with 102 of them receiving citations from 88 nodes within the cluster.

Each of these clusters can be considered `edge-cases', and while they are generally not very common, they need to be detected and either removed from the final set of author clusters forming  putative communities of practice (or reconfigured with other clusters to represent a different  community). 
Thus, further examination and possibly refinement of selected MCL clusters using citation data should be performed. 
 
Finally, while conductance and coherence have both been proposed as quality measures for evaluating clusters, they are not directly relevant to thematic relatedness, 
which requires human expert evaluation for reliable assessment. In our study,  we used human experts to evaluate  100 MCL clusters, which confirmed that the clusters 
with low conductance tended to have strong thematic relatedness.
However,  as this expert evaluation was limited to only 100 MCL clusters, and each
cluster was only evaluated by one reviewer, it is premature to draw definitive conclusions about the 
thematic relatedness of MCL clusters produced by this pipeline.

For each of the 35 MCL clusters produced by our pipeline, we  examined the authors of the publications in the cluster. The 35 publication clusters varied from 40--325 publications, with 199--1628 authors per cluster. 
The vast majority (between 70--91\%) of the authors contribute to only one paper each in a cluster. However, the 95th percentile of authors published from 2--4 publications in the relevant cluster, while the 99th percentile  
of authors publish from 2--10 publications per cluster. Thus, generally most authors participate only in minor ways in these clusters, but there are a few authors who are much more involved, a finding that is reminiscent of \cite{price_1966}.
Interestingly, the pattern of citation between authors in the 95th percentile varies from 0 to 100\% among the clusters, suggesting considerable heterogeneity in terms of collaboration and citation practice at
the upper end of productivity in these communities. 

\begin{table}
\hspace*{-1cm}
\scalebox{0.8}{
\begin{tabular}{rcrrrrrccccc}
  \hline
row & match\_year & size(m) & size(g) & cond(m) & cond(g) & coh(m) & coh(g) & int.edges(m) & int.edges(g) & jc \\ 
  \hline
1 & 1988 & 190 & 194 & 0.06 & 1.00 & 0.11 & 0.11 & 189 & 0 & 0.96\Tstrut \\ 
2 & 1990 & 328 & 327 & 0.01 & 1.00 & 0.08 & 0.08 & 327 & 0 & 0.99 \\ 
3 & 1993 & 114 & 112 & 0.01 & 0.01 & 0.11 & 0.11 & 113 & 111 & 0.98 \\ 
4 & 1988 & 211 & 224 & 0.04 & 1.00 & 0.10 & 0.09 & 210 & 0 & 0.92\\ 
5 & 1986 & 185 & 183 & 0.02 & 1.00 & 0.04 & 0.04 & 184 & 0 & 0.98 \\ 
6 & 1989 & 182 & 184 & 0.05 & 1.00 & 0.05 & 0.05 & 181 & 0 & 0.97 \\ 
7 & 1993 & 173 & 171 & 0.01 & 1.00 & 0.10 & 0.10 & 172 & 0  & 0.98 \\ 
8 & 1995 & 201 & 211 & 0.06 & 1.00 & 0.09 & 0.09 & 200 & 0  & 0.91 \\ 
9 & 1985 & 225 & 225 & 0.01 & 1.00 & 0.10 & 0.10 & 224 & 0  & 0.98 \\ 
10 & 1986 & 184 & 188 & 0.02 & 1.00 & 0.10 & 0.10 & 183 & 0  & 0.96 \\ 
11 & 1986 & 167 & 165 & 0.00 & 1.00 & 0.13 & 0.13 & 166 & 0  & 0.99 \\ 
12 & 1990 & 146 & 147 & 0.02 & 1.00 & 0.10 & 0.10 & 145 & 0  & 0.97 \\ 
13 & 1995 & 165 & 168 & 0.02 & 1.00 & 0.11 & 0.11 & 164 & 0  & 0.96 \\ 
14 & 1993 & 168 & 166 & 0.03 & 1.00 & 0.13 & 0.12 & 167 & 0  & 0.98 \\ 
15 & 1992 & 153 & 156 & 0.05 & 1.00 & 0.08 & 0.08 & 152 & 0  & 0.94 \\ 
16 & 1991 & 155 & 150 & 0.17 & 1.00 & 0.12 & 0.12 & 154 & 0 & 0.96 \\ 
17 & 1985 & 213 & 217 & 0.08 & 1.00 & 0.09 & 0.09 & 212 & 0 & 0.95 \\ 
18 & 1987 & 150 & 156 & 0.03 & 1.00 & 0.08 & 0.08 & 149 & 0  & 0.94 \\ 
19 & 1993 & 121 & 121 & 0.02 & 0.02 & 0.07 & 0.07 & 120 & 120 & 1.00 \\ 
20 & 1987 & 151 & 153 & 0.04 & 1.00 & 0.08 & 0.07 & 150 & 0  & 0.96 \\ 
21 & 1994 & 175 & 171 & 0.09 & 1.00 & 0.13 & 0.13 & 174 & 0  & 0.93 \\ 
22 & 1991 & 141 & 150 & 0.12 & 1.00 & 0.06 & 0.06 & 140 & 0  & 0.90 \\ 
23 & 1994 & 328 & 332 & 0.05 & 1.00 & 0.14 & 0.14 & 327 & 0  & 0.95 \\ 
24 & 1988 & 264 & 272 & 0.05 & 1.00 & 0.09 & 0.09 & 263 & 0  & 0.95 \\ 
25 & 1988 & 139 & 137 & 0.02 & 1.00 & 0.05 & 0.05 & 138 & 0  & 0.97 \\ 
26 & 1988 & 156 & 159 & 0.03 & 1.00 & 0.10 & 0.10 & 155 & 0  & 0.96 \\ 
27 & 1994 & 128 & 124 & 0.10 & 1.00 & 0.08 & 0.08 & 127 & 0  & 0.97 \\ 
28 & 1994 & 116 & 109 & 0.15 & 0.15 & 0.07 & 0.07 & 115 & 108  & 0.92 \\ 
29 & 1993 & 106 & 106 & 0.01 & 0.01 & 0.02 & 0.02 & 175 & 175  & 1.00 \\ 
30 & 1993 & 118 & 113 & 0.06 & 0.06 & 0.11 & 0.11 & 118 & 112  & 0.93 \\ 
31 & 1994 &  97 &  97 & 0.01 & 0.01 & 0.08 & 0.08 & 96 & 96  & 1.00 \\ 
32 & 1992 &  78 &  78 & 0.00 & 0.00 & 0.13 & 0.13 & 77 & 77  & 1.00 \\ 
33 & 1990 &  55 &  56 & 0.02 & 0.02 & 0.08 & 0.08 & 54 & 55  & 0.98 \\ 
34 & 1987 &  47 &  47 & 0.02 & 0.02 & 0.08 & 0.08 & 46 & 46  & 1.00 \\ 
35 & 1990 &  68 &  66 & 0.08 & 0.08 & 0.08 & 0.08 & 67 & 65  & 0.94 \\ 
   \hline
\end{tabular}}
\caption{Features of the 35 MCL-Graclus pairs of clusters selected for thematic relatedness. \textbf{m,g} refer to MCL and Graclus respectively. size: number of publications in the cluster. cond: conductance, coh: coherence, int.edges: number of internal edges, jc: Jaccard Coefficient
for publications in paired MCL and Graclus clusters. All 35 MCL clusters received the top rating (1) by the evaluators and the Jaccard Coefficient for MCL-Graclus pairs is greater than 0.9 for all pairs (row 22 shows Jaccard Coeffficient of 0.90, but that is due to rounding). The median conductance of MCL is 0.032, the median conductance of the Graclus clusters is 1.0 (25 Graclus clusters have a conductance of 1.0).
Rows 1, 2, and 3 are discussed further in the text.
}
\label{tab:tab3}
\end{table}

Using Scopus author identifiers, we also examined the distribution of the 2,822,497 unique authors who contributed to the set of 134,094 clusters to ask how many clusters each belonged to. Roughly 61\% authored documents 
in only one cluster and 89\% authored documents in 5 clusters or less with an average of 3.1 clusters per author. However, we discovered several authors associated with very large numbers of clusters (e.g., one author had  1,005 papers 
distributed across 623 clusters) and  0.08\% of 2,822,497 authors contributed to 100 clusters or more. However, there are different contexts that can result in large values like these.
For example, an author may  write many papers that cite papers in different clusters, and a different  author may write several papers that are in different clusters, each of which is cited substantially, and 
both authors will have papers from many clusters.
To understand these different scenarios, it is necessary to examine in-graph and in-Scopus citation counts: comparing the number of citations received versus 
those made, therefore, enables the discovery of authors of influence (cited) versus those who cite influential papers without receiving many (or any) citations themselves. In the case of the author with 1,005 publications 
in 623 clusters, this person authored over 2,000 publications indexed in Scopus (confirmed independently in a search of PubMed) and cited 911 articles of which 163 were self-citations. In addition these 
1,005 articles were cited 193 times (Supplementary Data).

\section*{Conclusions} 

This study reveals some interesting trends that suggest directions for future work. 
First,  our research suggests a pipeline to identify possible communities of practice: (1) use convergent clustering, in this both Markov Clustering (MCL) and Graclus (a spectral clustering method), to  cluster the citation network, (2) find those MCL clusters within an appropriate size range that have very low conductance values and high Jaccard coefficients to their matched Graclus clusters, and (3) extract the author community for each article cluster, and  filter out those communities that are edge cases (e.g., clusters where only one author is citing the other authors, or where only one author is cited by any other author).

Although we were able to conduct an expert evaluation for only a very small sample of the clusters produced by this pipeline, our experts ranked these clusters  highly  with respect to thematic relatedness, suggesting that this pipeline could produce article clusters derived from a scientific theme. As a result, the author communities detected using the pipeline represent likely communities of practice, as was our objective.

We applied a stringent Jaccard coefficient to the article clusters we considered in order to reduce the false discovery rate (FDR). However, the conductance values even for lower Jaccard coefficients also resulted in comparably low median conductance when we examined the 100 clusters evaluated by humans. For example, the median conductance of MCL clusters was 0.03 when the Jaccard coefficient was greater than 0.9 and 0.08 when the Jaccard coefficient was relaxed to greater than 0.7. These data suggest that relaxing these constraints should be rigorously explored since it may not increase the FDR significantly. This hypothesis is also supported by the high ratings that the experts gave to MCL clusters that had low conductance but lower overlap with their paired Graclus clusters. 

Despite these promising results, we are well aware of the limitations of using citation and cluster analysis to identify communities of practice. The best techniques would ideally use expert evaluation, which is unfortunately not scalable (and in our study, we only used expert evaluation for 100 of the clusters we generated). Furthermore, this study only examined immunology publications and others connected to them by citation. Other studies have shown that citation behavior can depend significantly on the field, making extrapolation of trends from one field to another premature \citep{Wallace2012,Bradley2020}. Thus, the trends in this study may not be consistently found in other research disciplines or timeframes. Our future work will characterize these initial observations, evaluate additional clustering techniques, and focus on elucidating interactions between authors within and across clusters to refine the pipeline we envision.
\clearpage

\section{Supportive Information} \vspace{3mm}Supplementary material  and code used in this study is available on our Github site~\citep{Korobskiy2019}.

\subsection*{Acknowledgments}
\vspace{3mm} We thank Vladimir Smirnov for helpful discussions on using Markov Clustering. The ERNIE project involves a collaboration with Elsevier. The content of this publication is solely the responsibility of the authors and does not necessarily represent the official views of the National Institutes of Health or Elsevier.  We thank our Elsevier colleagues for their support of the ERNIE project.

\subsection*{authorcontributions} 
\vspace{3mm}
Shreya Chandrasekharan: Conceptualization; Methodology; Investigation; Writing—Original Draft; Writing—Review and Editing. Mariam Zaka: Investigation; Writing—Review and Editing.; Stephen Gallo: Investigation; Writing—Review and Editing.Tandy Warnow: Conceptualization; Methodology; Writing—Original Draft; Writing—Review and Editing. George Chacko: Conceptualization; Methodology; Investigation; Writing—Original Draft; Writing—Review and Editing; Funding Acquisition, Resources; Supervision.

\section{Competing Interests} \vspace{3mm} The authors have no competing interests. Scopus data used in this study was available to us through a collaborative agreement with Elsevier on the ERNIE project. Elsevier personnel played no role in conceptualization, experimental design, review of results, or conclusions presented. 

\section{Funding Information} \vspace{3mm} Research and development reported in this publication was partially supported by federal funds from the National Institute on Drug Abuse (NIDA), National Institutes of Health, U.S. Department of Health and Human Services, under Contract Nos. HHSN271201700053C (N43DA-17-1216) and HHSN271201800040C (N44DA-18-1216). Tandy Warnow receives funding from the Grainger Foundation.

\section{Data Availability} \vspace{3mm} Access to the bibliographic data analyzed in this study requires a license from Elsevier. Code generated for this study is freely available from our Github site~\citep{Korobskiy2019}.

\newpage

\bibliography{tpa}

\newpage
\section {Figures}

\newpage

\section{Supportive Information} \vspace{3mm}Supplementary material is available on our Github site~\citep{Korobskiy2019}.

\end{document}